\documentclass[a4paper,12pt]{article}
\usepackage[utf8]{inputenc}
\usepackage{amsmath,amssymb,amsfonts,amsthm}

\usepackage{graphicx}
\usepackage{wasysym}
\usepackage{xcolor}

\usepackage{subcaption}

\title{Spherically symmetric linear perturbations of electrically counterpoised dust}
\author{Andrés Aceña and Ivan Gentile de Austria\\
\\
Facultad de Ciencias Exactas y Naturales,\\
  Universidad Nacional de Cuyo, \\ Instituto Interdisciplinario de Ciencias B\'asicas, CONICET, \\ Padre Jorge Contreras 1300, M5502JMA Mendoza, Argentina.}

\begin{document}

\maketitle

\begin{abstract}
    We consider spherically symmetric linear perturbations of static spherically symmetric spacetimes where the matter content is ellectrically counterpoised dust. We show that the evolution equation for the fluid perturbation implies that the fluid elements move with constant velocities. Therefore there are neither oscillations nor exponential departure from the background solution. We present an explicit example that shows that the perturbation could lead to the formation of a black hole.
\end{abstract}

\section{Introduction}

Extracting general conclusions from the Einstein field equations has always been a difficult task, specially if matter is included. A particularly simple but useful matter model is electrically counterpoised dust (ECD). It corresponds to a charged perfect fluid without pressure. Although it might seem that including charge, therefore having to deal with the Einstein-Maxwell system of equations, would make everything more complicated, it can be seen that if the matter and charge density are perfectly balanced, then any static distribution of matter is a solution of the field equations. This was first noticed by Majumdar \cite{Majumdar1947} and Papapetrou \cite{Papapetrou1947} for a system consisting of discrete particles. If one considers only electrovacuum, then to each particle there is an event horizon that is interpreted as an extremal Reissner-Nordstr\"om (ERN) black hole \cite{Hartle1972}. One way of completing the Majumdar-Papapetrou spacetimes is by matching the exterior solution to static interior solutions made of ECD, as shown in \cite{Das1962}, and more recently analyzed in detail in \cite{Varela2003}. Also, the Majumdar-Papapetrou formalism can be extended to higher dimensions, keeping its distinguishing features \cite{Lemos2005}.

This extremely particular property of ECD, that any static charge distribution gives rise to a solution of the Einstein-Maxwell field equations, has been exploited to construct spacetimes with particular characteristics. In this way, features that proved difficult for a general analysis, allowed an analytic treatment for particular solutions. Examples of this includes the study of the relation between charge and mass in the Reissner-Nordstr\"om spacetimes and the construction of a point charge model \cite{Bonnor1960}. Also, in \cite{Bonnor1972}, ECD is used to construct static objects with unbounded density, and in \cite{Bonnor1975} it is shown that redshifts as high as desired can be obtained from regular objects. Even more elaborate analytic models can be constructed, in particular ECD spheroids, that allow to discuss the hoop conjecture \cite{Bonnor1998}. A special characteristic of the constructed solutions is that they can be made to be close to the ERN black hole. This has been studied in connection with the bifurcation of the solutions \cite{Horvat2005}, and in \cite{Meinel2011} it has been shown that such black hole limit is a general feature of ECD solutions.

Although extremely interesting for analyzing such difficult general questions, one underlying assumption when generalizing particular results of explicit solutions is that said solutions are stable. If the solution is stable one expects that physical realistic situations that are close to the solution could appear in nature, although the exact solution would not, as it would always be subjected to some perturbations. On the other hand, if the solution is unstable, then there is no hope of finding it in nature, and the general conclusions that have been extracted lose ground. The discussion of ECD regarding stability is, as in general with stability concerns, not an easy one. Leaving aside that dust would always be an approximation where thermal energy has been discarded, the fact that the mass and charge density needs to be equal implies a particular fine tuning difficult to justify from more fundamental matter models. No known particle satisfy such relationship, and in fact, it is generally grossly violated. For example, if one wants to make such a matter system starting with neutral hydrogen, one needs to ionize only one in $10^{18}$ atoms. Such is the disparity between electric and gravitational force for fundamental particles. As ECD has the same mass and charge density, then it is considered to be extremal in tis mass-charge relationship, and corresponds microscopically to the ERN black hole. This is also a reflection of the fact that from the perspective of general relativity all fundamental particles are over-extremal.

Another point that calls into question the stability of static ECD solutions is their Newtonian counterparts. If we single out a portion of the fluid, then the total force on such portion vanish. If it starts to move then the restitutive force must come from gravitational and electromagnetic interactions, as there is no pressure, and therefore it has to be of second order in the perturbation. This implies that the acceleration would be of second order on the perturbation. The argument in itself does not mean that the system is unstable, but shows that the perturbations have a grater chance of growing.

Closely related to the question of stability of ECD spacetimes is the stability of charged fluid spheres with pressure. This problem was tackled in \cite{Anninos2001}, where the Tolman-Oppenheimer-Volkov equations were integrated numerically for several equations of state and the stability was examined by both a normal mode and an
energy analysis. It was found that in general there is a stability limit, beyond which the spheres are unstable and therefore undergo gravitational collapse, in all cases before they reach the ERN limit $Q=M$.

In this article we take a first step towards determining the stability of ECD spacetimes. We consider static spherically symmetric solutions of the Einstein-Maxwell system of equations with ECD as the matter model, and analyze the behaviour of linear spherically symmetric perturbations. The article is organized as follows. In Section \ref{secSpacetime} we present the equations that the ECD spacetime needs to satisfy. Then the perturbations are analyzed in Section \ref{secPert}. In Section \ref{secExample} an explicit example is constructed, which shows that if the original spacetime is close to a black hole, then the perturbation can make the spacetime to collapse into a black hole. The conclusions are discussed in Section \ref{secConc}.

\section{Spherically symmetric ECD spacetime}\label{secSpacetime}

The contents of this section are well known and the reader can refer for example to \cite{Bonnor1960} and \cite{Anninos2001} . The Einstein-Maxwell system of equations is
\begin{equation}
 R_{\mu\nu}-\frac{1}{2}g_{\mu\nu}R=8\pi(T^{d}_{\mu\nu}+T^{e}_{\mu\nu}),
\end{equation}
\begin{equation}
 \nabla_{\alpha}F^{\mu\alpha}=4\pi j^{\mu},
\end{equation}
where the electromagnetic tensor is given in terms of the electromagnetic potential
\begin{equation}
 F_{\mu\nu}=\partial_{\mu}A_{\nu}-\partial_{\nu}A_{\mu},
\end{equation}
and the associated energy-momentum tensor is
\begin{equation}
 T^e_{\mu\nu} = \frac{1}{4\pi} \left(F_{\alpha\mu}F^\alpha\,_\nu-\frac{1}{4}F_{\alpha\beta}F^{\alpha\beta}g_{\mu\nu}\right).
\end{equation}
The matter model is dust, the corresponding energy-momentum tensor being
\begin{equation}
 T^{d}_{\mu\nu}=\rho\, u_{\mu}u_{\nu},
\end{equation}
where $u^\mu$ is the four-velocity of the fluid and $\rho$ is the mass density. The current density is given in terms of the charge density, $\sigma$, by
\begin{equation}
j^{\mu}=\sigma u^{\mu}.
\end{equation}

Let us consider a static spherically symmetric spacetime, whose metric can always be written in Schwarzschild coordinates as
\begin{equation}\label{eqMetric}
    ds^2 = -e^{2\Phi(r)}dt^2+e^{2\Lambda(r)}dr^2+r^2(d\theta^2+\sin^2\theta d\phi^2).
\end{equation}
The four-velocity of the fluid is
\begin{equation}
    u^{\mu} = e^{-\Phi(r)}\,\partial_t,
\end{equation}
and the electromagnetic potential can be expressed in terms of a scalar potential $\nu(r)$,
\begin{equation}
    A_\mu = \nu(r)\, dt.
\end{equation}
Denoting by prime the derivative with respect to $r$, the field equations are
\begin{equation}\label{eqLambda}
    e^\Lambda = 1 + r\Phi',
\end{equation}
\begin{equation}\label{eqRho}
    \rho = \frac{\Phi'' + \Phi'^2 + 2\Phi'/r}{4\pi(1+r\Phi')^3},
\end{equation}
\begin{equation}
    \nu = e^\Phi,
\end{equation}
and
\begin{equation}\label{eqSigma}
    \sigma = k\rho,
\end{equation}
with $k=\pm 1$ (in the following we take $k=1$, as it simply ammounts to the convention of which charge is considered positive).

The usual procedure to obtain a solution that represents a compact object is to consider a ball of coordinate radius $r=R$, in whose interior the mass (and correspondingly charge) density is different from zero, and outside there is electrovacuum.
Then, the exterior solution is ERN,
\begin{equation}
    \Phi_E(r) = - \Lambda_E(r) = \ln \left(1-\frac{M}{r}\right),
\end{equation}
where $M$ is the ADM mass of the spacetime and the subindex $E$ means that the quantities refer to the exterior region ($r>R$). For the interior region ($r<R$) we leave the functions without any subindex. 
To obtain an interior solution we can follow the typical procedure of making a simple ansatz for the metric function $\Phi$. Then we glue the interior and exterior functions using the juncture conditions, which are
\begin{equation}
    \Phi(R) = \Phi_E(R),\quad \Phi'(R) = \Phi_E'(R),\quad \Lambda(R) = \Lambda_E(R),\quad \Lambda'(R) = \Lambda_E'(R).
\end{equation}
Also, to ensure regularity at $r=0$
\begin{equation}
    \Lambda(0) = 0,\quad \Lambda'(0)=0,\quad \Phi'(0) = 0.
\end{equation}
Since an ansatz for $\Phi$ is made, all the previous conditions are satisfied if we enforce that
\begin{equation}
    \Phi(0)=0,\quad \Phi(R) = \Phi_E(R),\quad \Phi'(R) = \Phi_E'(R),\quad \Phi''(R) = \Phi_E''(R).
\end{equation}
The function $\Lambda$ and the mass density $\rho$ are then calculated using \eqref{eqLambda} and \eqref{eqRho}.

For the discussion of Section \ref{secExample} it is convenient to use the charge inside a sphere of radius $r$ as the function to make the ansatz. The charge can be integrated explicitly using the equations \eqref{eqSigma}, \eqref{eqRho} and \eqref{eqLambda}, as
\begin{equation}
    Q = \int_V \sigma\,dV = 4\pi\int_0^r r^2\,e^\Lambda\,\sigma\,dr = r(1-e^{-\Lambda}).
\end{equation}
Now we can write the other functions in terms of $Q$,
\begin{equation}
    \Lambda = -\ln\left(1-\frac{Q}{r}\right),
\end{equation}
\begin{equation}\label{eqPhi}
    \Phi' = \frac{Q}{r(r-Q)},
\end{equation}
\begin{equation}
    \rho = \frac{Q'(r-Q)}{4\pi r^3}.
\end{equation}
The regularity and junction conditions then become
\begin{equation}
    Q(0)=0,\quad Q'(0)=0,\quad Q''(0)=0,\quad Q(R)=M,\quad Q'(R)=0,
\end{equation}
and the integration constant in \eqref{eqPhi} is fixed by $\Phi(R) = \Phi_E(R)$.
From the previous equations we see that if $Q=r$ for any $0<r\leq R$ then the solution fails to be regular. This is expected because if $Q=r$ then there is an event horizon and the solution is no longer a regular ECD spacetime but the ERN black hole. We will use the difference between $r$ and $Q$ in Section \ref{secExample} as a measure of how far the solution is from forming a black hole.

\section{Spherically symmetric linear perturbations}\label{secPert}

Given the symmetries of the solution, we study the simplest possible perturbations, the linear spherical ones. We follow \cite{Anninos2001} and here we omit the lengthy steps to obtain the first order equations, as the procedure is completely analogous. As in Section \ref{secSpacetime}, the metric can again be written in the form \eqref{eqMetric}, but now the functions $\Phi$ and $\Lambda$ depend also on $t$. As we consider perturbations, these functions are written as
\begin{equation}
 \label{20}
 \Phi(t,r)=\Phi_{0}(r)+\Phi_{1}(t,r),\quad\Lambda(t,r)=\Lambda_{0}(r)+\Lambda_{1}(t,r),
\end{equation}
where the subindex $0$ is used to denote the unperturbed background quantities and the subindex $1$ for the perturbation, which is assumed to be small with respect to the unperturbed quantity, and with the corresponding expressions for the other functions.
The fundamental quantity in the perturbation scheme is the displacement of a fluid element.
A fluid element located at coordinate radius $r$ in the unperturbed configuration is displaced to coordinate radius $r + \xi(r,t)$ at coordinate time $t$ in the perturbed configuration. 

As it turns out, the first order perturbation of all the quantities are functions of the unperturbed quantities and of $\xi$. We have for these quantities
\begin{equation}
    \Lambda_1=-4\pi r e^{2\Lambda_0}\rho_0 \xi,
\end{equation}
\begin{equation}
    \Phi_1' = -4\pi e^{3\Lambda_0}\rho_0 \xi,
\end{equation}
\begin{equation}
    \rho_1 = -\rho_0 \xi' - \left[\rho_0'+\left(\frac{2}{r}-\Phi_0'\right)\rho_0\right]\xi,
\end{equation}
\begin{equation}
    Q_1 = -Q_0'\xi.
\end{equation}
The equation for the fluid displacement turns out to be simply
\begin{equation}\label{eqXi}
    \frac{\partial^2\xi}{\partial t^2} = 0.
\end{equation}
Therefore we can freely specify two functions of $r$, $\xi_0(r)$ and $\xi_1(r)$, and the solution to \eqref{eqXi} is
\begin{equation}
    \xi(r,t) = \xi_0(r) + \xi_1(r)\,t.
\end{equation}
The function $\xi_0(r)$ corresponds to the initial displacement of the fluid, while $\xi_1(r)$ encodes the velocity of the fluid element.
The boundary conditions are
\begin{equation}
    \lim_{r\rightarrow 0}\frac{\xi}{r} = constant,\quad \lim_{r\rightarrow R}\xi=0.
\end{equation}
The first condition is imposed in order to have a regular behavior of the mass density at the origin, while the second condition ensures that a surface density mass is not generated at the boundary of the object.

For the discussion in the following section it is convenient to work with adimensional quantities. We define
\begin{equation}
    x:=\frac{r}{R},\quad \mu:=\frac{M}{R},\quad q:=\frac{Q}{R},\quad \zeta:=\frac{\xi}{R}, \quad \tilde{t} := \frac{t}{R}.
\end{equation}
Then the interior correspond to $0<x<1$ and we write the solution in terms of $q_0$. Denoting by a dot the derivative with respect to $x$ we have
\begin{equation}
    \Lambda_0 = -\ln\left(1-\frac{q_0}{x}\right),
\end{equation}
\begin{equation}
    \dot{\Phi}_0 = \frac{q_0}{x(x-q_0)},
\end{equation}
and for the perturbation
\begin{equation}
\Lambda_1 = -\frac{\dot{q}_0}{x-q_0}\zeta,
\end{equation}
\begin{equation}
    \dot{\Phi}_1 = -\frac{\dot{q}_0}{(x-q_0)^2}\zeta,
\end{equation}
\begin{equation}
    q_1 = -\dot{q}_0 \zeta,
\end{equation}
with
\begin{equation}\label{eqZeta}
    \zeta(x,t) = \zeta_0(x) + \zeta_1(x)\tilde{t},
\end{equation}
and where $\zeta_0$ and $\zeta_1$ are arbitrary functions of $x$ satisfying
\begin{equation}
    \lim_{x\rightarrow 0}\frac{\zeta}{x} = constant,\quad \lim_{x\rightarrow 1}\zeta=0.
\end{equation}
Also, the function $q_0$ needs to satisfy
\begin{equation}\label{condQ}
    q_0(0) = 0, \quad \dot{q}_0(0) = 0, \quad \ddot{q}_0(0) = 0, \quad q_0(1) = \mu,\quad \dot{q}_0(1) = 0.
\end{equation}

\section{An explicit example}\label{secExample}

In this section we present as explicit example a family of ECD spacetimes, and consider the linear perturbations on them. The purpose of this section is to show that if the original spacetime is close to forming a black hole, then by a small perturbation a black hole can be formed.

\begin{figure}
 \centering
 \begin{subfigure}{.5\textwidth}
   \centering
   \includegraphics[width=\linewidth]{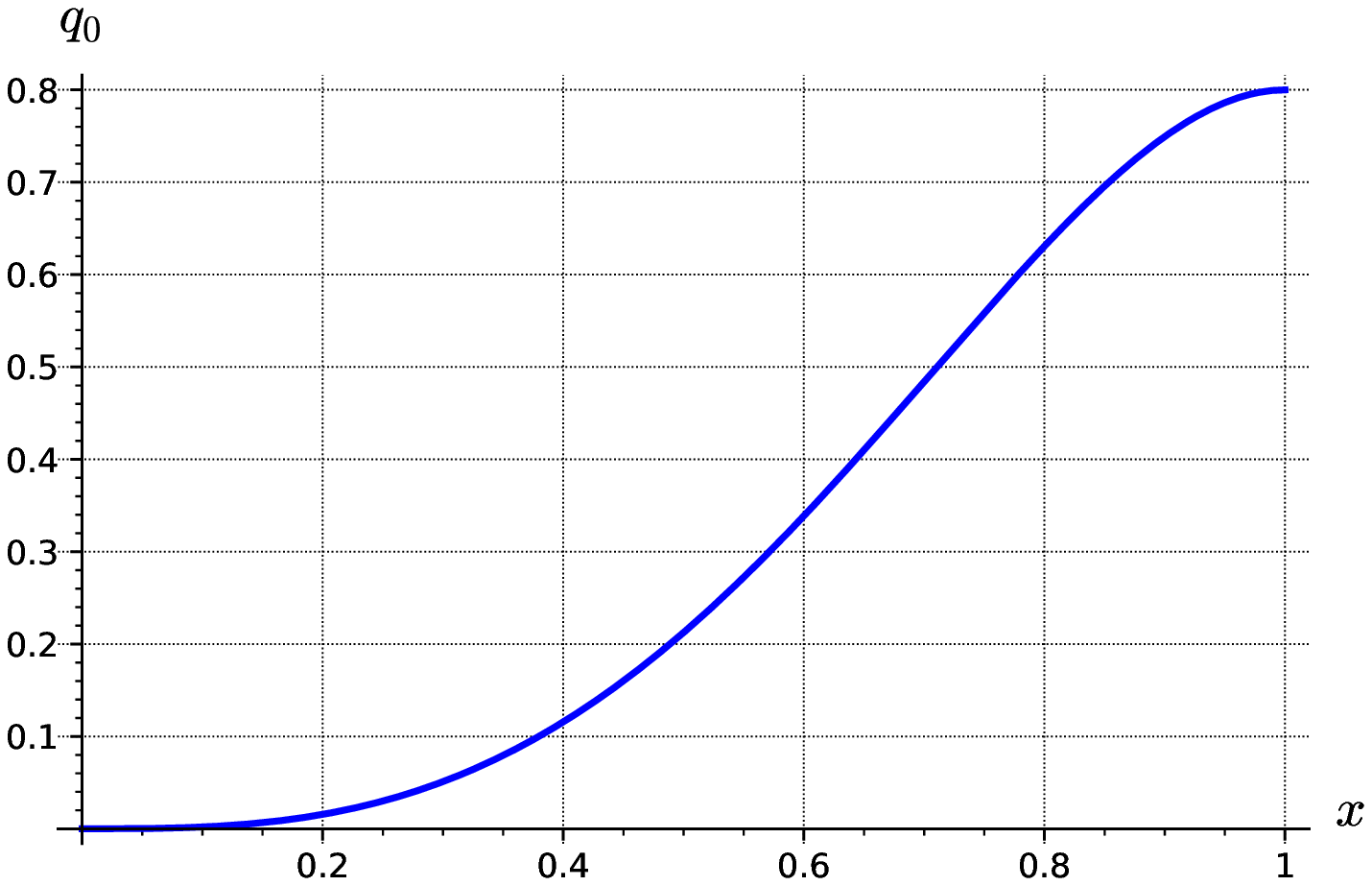}
 \end{subfigure}%
 \begin{subfigure}{.5\textwidth}
   \centering
   \includegraphics[width=\linewidth]{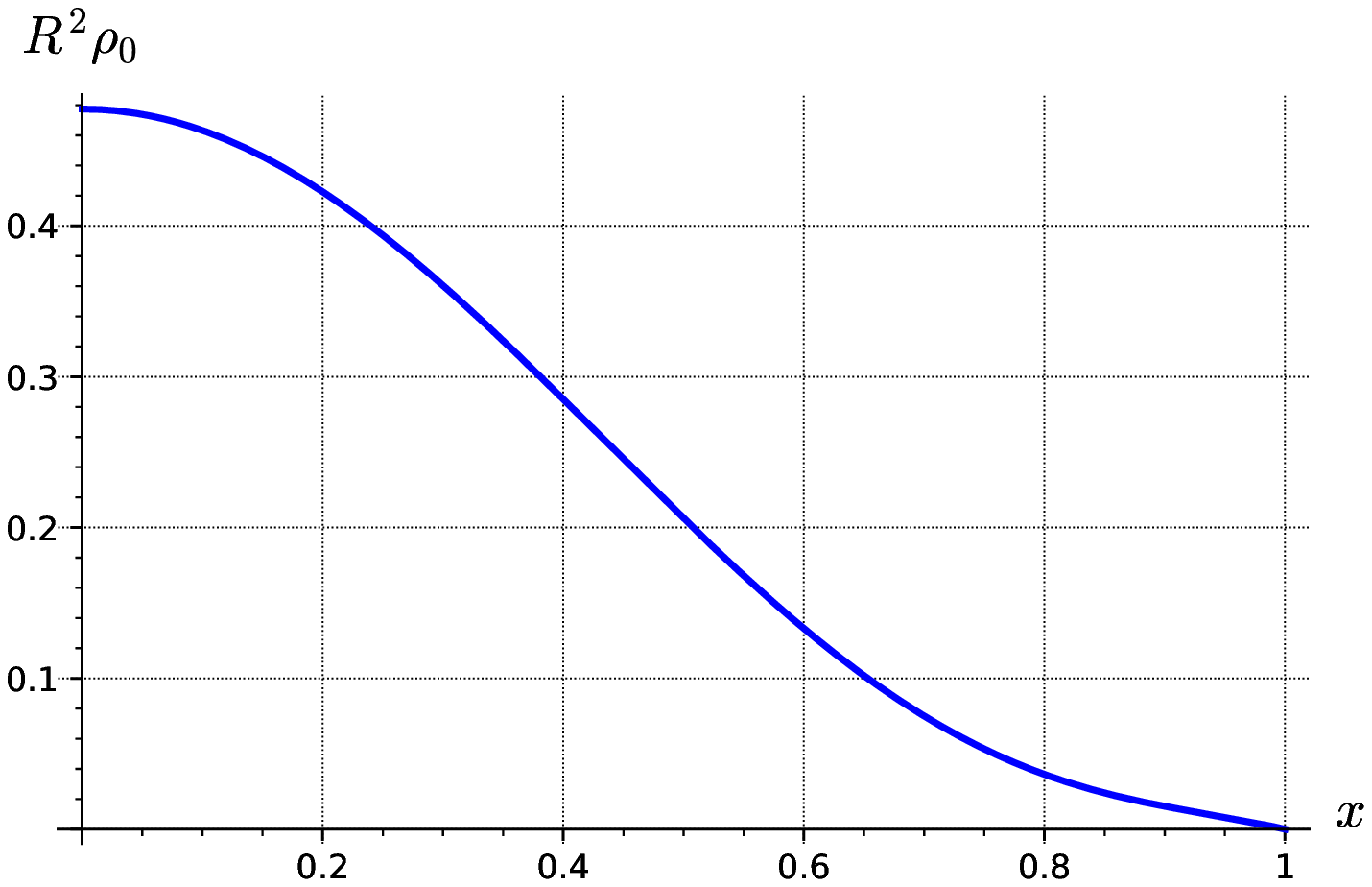}
 \end{subfigure}
 \caption{Carga y densidad de carga para $\mu=\frac{4}{5}$.}
 \label{fig1}
 \end{figure}

\begin{figure}
 \centering
 \begin{subfigure}{.5\textwidth}
   \centering
   \includegraphics[width=\linewidth]{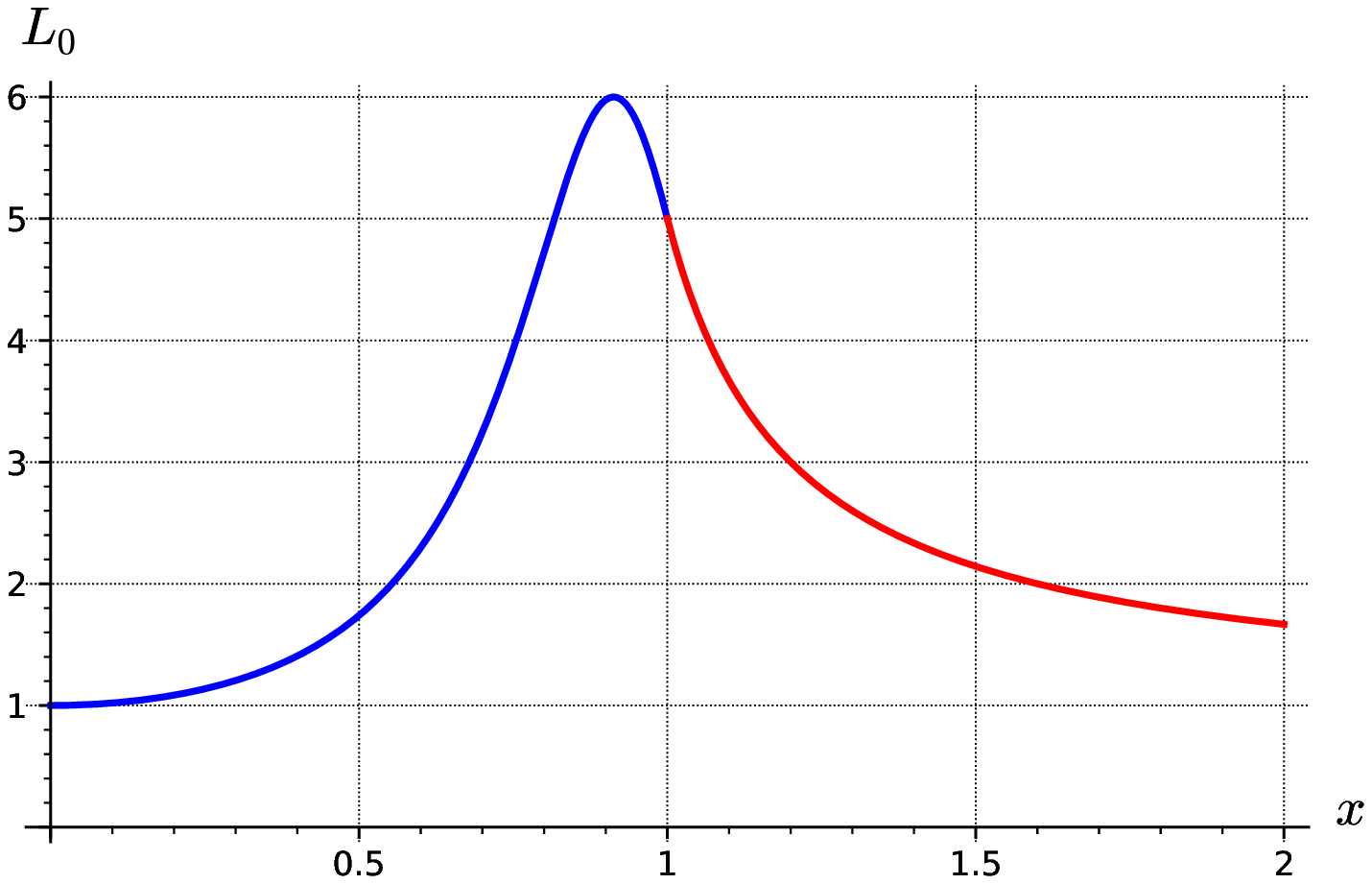}
 \end{subfigure}%
 \begin{subfigure}{.5\textwidth}
   \centering
   \includegraphics[width=\linewidth]{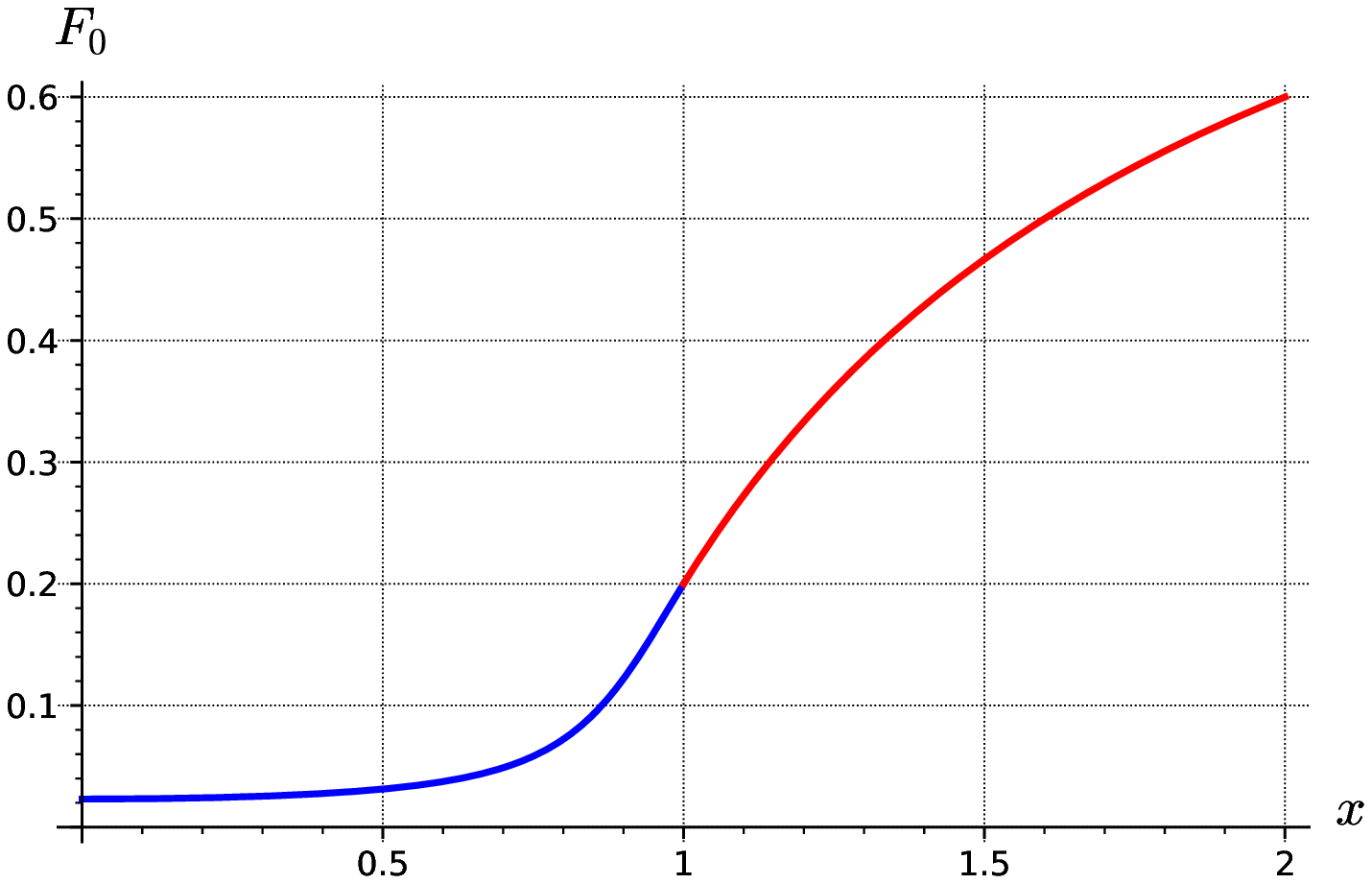}
 \end{subfigure}
 \caption{Funciones $L_0$ y $F_0$ para $\mu=\frac{4}{5}$.}
 \label{fig2}
 \end{figure}

We consider the parameter $\mu$, that correspond to the total mass and charge of the spacetime, to parametrize the family of solutions. We choose the function $q_0$ simply as a polynomial in $x$ that satisfy \eqref{condQ},
\begin{equation}
    q_0 = \frac{\mu}{2}x^3(5-3x^2).
\end{equation}
Then the functions $\Lambda_0$ and $\Phi_0$ can be explicitly calculated. For the discussion is convenient to use as metric functions
\begin{equation}
    L_0 := e^{\Lambda_0},\quad F_0 := e^{\Phi_0},
\end{equation}
and then
\begin{equation}
    L_0 = \frac{2}{2-\mu x^2(5-3x^2)},
\end{equation}
\begin{eqnarray}
    F_0  & = &  \frac{(1-\mu)^\frac{5}{4}}{\left[1-\frac{1}{2}\mu x^2(5-3x^2)\right]^\frac{1}{4}} \\
    & & \times \exp\left[-\frac{5\mu}{2\sqrt{24-25\mu}}\arctan\left(\frac{\sqrt{24-25\mu}(1-x^2)}{4-\mu(5-x^2)}\right)\right]
\end{eqnarray}
We see that if
\begin{equation}
    \mu \geq \frac{24}{25}.
\end{equation}
then $L_0$ diverges at
\begin{equation}
    x = \sqrt{\frac{5}{6}}\sqrt{1\pm\sqrt{1-\frac{24}{25\mu}}}.
\end{equation}
If $\mu<\frac{24}{25}$ the solution is regular for all values of $x$. In Figure \ref{fig1} we have plotted the function $q_0$ and the charge density for the spacetime with $\mu=\frac{4}{5}$. Also, the functions $L_0$ and $F_0$ are shown in Figure \ref{fig2}.

In order to see how close the spacetime is to forming a black hole we consider the function $q_0/x$. The maximum is attained at
\begin{equation}
    x_M = \sqrt{\frac{5}{6}},
\end{equation}
where its value is
\begin{equation}
    \frac{q_0(x_M)}{x_M} = \frac{25}{24}\mu.
\end{equation}
If $q/x=1$ we have a black hole and therefore we define the parameter
\begin{equation}
    \delta = 1 - \frac{25}{24}\mu
\end{equation}
as a measure of how close the regular spacetime is to a black hole spacetime.

Now we consider the linear perturbations on these spacetimes. The function $q$ to first order is
\begin{equation}
    q = q_0 + q_1 = q_0 - \dot{q}_0 \zeta,
\end{equation}
and
\begin{equation}
    \dot{q}_0 = \frac{15}{2}\mu x^2(1-x^2).
\end{equation}
Therefore we have that if
\begin{equation}
    \zeta(x_M,\tilde{t}) = -\sqrt{\frac{5}{6}}\frac{\delta}{1-\delta} \Rightarrow \frac{q(x_M)}{x_M} = 1.
\end{equation}
Here the exact value of $\zeta$ is of no real importance, what matters is its order of magnitude. Also, from \eqref{eqZeta} we see that if we choose $\zeta_1(x)=0$, just by choosing $\zeta_0(x)$ with negative sign and of order $\delta$ we can make a black hole. On the other hand, and more interesting, as it departs from staticity, is to choose $\zeta_0(x)=0$ and $\zeta_1(x)< 0$. This amounts to perturb the solution by giving the fluid elements an inward velocity. Then in a time proportional to $\delta$ a black hole is formed.

\section{Conclusions}\label{secConc}

We have considered linear spherically symmetric perturbations on static spherically symmetric ECD spacetimes. We have obtained the equation for the evolution of the displacement of the fluid elements under the perturbation. This equation is interpreted as that there is no net force acting on the fluid elements, and therefore the displacement grows linearly in time. There is no restitutive force acting on the elements, but also the perturbation does not grow exponentially. This corresponds to the so called indifferent equilibrium of the solution. Although it has been argued that the solution is in such an equilibrium, and therefore it is stable, we consider that our treatment sheds light on the specifics of the ECD spacetimes and their equilibria. In particular, we constructed a family of solutions where there is a clear idea of how far the solution is from forming a black hole. We showed that in a time proportional to the distance to being a black hole the perturbation leads to a concentration of charge sufficient to form a black hole. Although the static solution is indeed indifferently stable, in general relativity this does not mean what one would expect from a stable spacetime. If the ECD solution is perturbed by giving a velocity distribution to the fluid elements then the perturbation grows linearly and never settles to an ECD solution, more so, it never returns to the original configuration. Also, in a more extreme scenario, the perturbation could lead to the formation of a black hole, changing the solution radically. In this sense we do not consider the ECD solutions to be stable, and we think that caution should be exercised when extracting conclusions for astronomical objects from such spacetimes, as it has been done for example with regard to possible redshifts.

\section*{Acknowledgements}
Several calculations were performed and the figures produced in SageMath \cite{SageMath} with the use of the package SageManifolds \cite{SageManifolds}.

This work was supported by grants PIP 112-201301-00532 of CONICET, Argentina, and M076 and M060 of SIIP, Universidad Nacional de Cuyo, Argentina.


\end{document}